\documentclass[11pt,journal]{IEEEtran}
\IEEEoverridecommandlockouts

\usepackage[noadjust]{cite}

\usepackage[cmex10]{amsmath}
\usepackage{algorithmic,amssymb,graphicx}

\begin{document}
\title{Utility of Beamforming Strategies for Secrecy in Multiuser MIMO Wiretap Channels}

\author{\authorblockN{Amitav Mukherjee and A. Lee Swindlehurst\\
\authorblockA{Dept.~of Electrical Engineering \& Computer Science\\
University of California\\
Irvine, CA 92697-2625}}}

\maketitle

\begin{abstract}
This paper examines linear beamforming methods for secure
communications in a multiuser wiretap channel with a single
transmitter, multiple legitimate receivers, and a single eavesdropper,
where all nodes are equipped with multiple antennas. No information
regarding the eavesdropper is presumed at the transmitter, and we
examine both the broadcast MIMO downlink with independent information,
and the multicast MIMO downlink with common information for all
legitimate receivers. In both cases the information signal is
transmitted with just enough power to guarantee a certain SINR at the
desired receivers, while the remainder of the power is used to
broadcast artificial noise. The artificial interference selectively
degrades the passive eavesdropper's signal while remaining orthogonal
to the desired receivers. We analyze the confidentiality provided by
zero-forcing and optimal minimum-power beamforming designs for the
broadcast channel, and optimal minimum-MSE beamformers for the
multicast channel. Numerical simulations for the
relative SINR and BER performance of the eavesdropper demonstrate the
effectiveness of the proposed physical-layer security schemes.
\end{abstract}

\section{INTRODUCTION}
The growing interest in security at the physical layer of wireless
communications has sparked a resurgence of research in
information-theoretic secrecy. Physical layer security incorporates
signal and code design to limit the information that can be extracted
by an eavesdropper at the bit level, as a supplement to classical
cryptographic security at the link or higher layers. Wyner's landmark
paper on secure communications in a point-to-point wiretap channel
\cite{Wyner75} paved the way for characterizing the secrecy capacity
of specific types of multiuser broadcast, interference, and
multiple-access channels with single-antenna nodes
\cite{Csizar78}--\cite{Yates08}, although their general secrecy
capacity regions under fading remain mostly unknown. Similarly, the
secrecy capacity achievable in multiple-input multiple-output (MIMO)
multiuser networks is largely an open problem, with limited results
available for MIMO broadcast channels with two receivers
\cite{Poor09}. In the majority of the literature on confidential
transmissions in multiuser networks, knowledge of the probability
distribution of the eavesdropper's channel is assumed at the
transmitter, which inherently provides information about the number of
eavesdropper antennas as well.

Motivated by the above, this paper studies the effectiveness of simple
beamforming strategies for maintaining confidentiality in a MIMO
downlink system with multiple legitimate multi-antenna receivers and a
single passive eavesdropper with an unknown channel distribution. A
portion of the transmit power is used to broadcast the information
signal vector with just enough power to guarantee a certain
sig\-nal-to-int\-er\-fer\-ence-plus-noise ratio (SINR) for
the intended receivers, and the remainder of the power is used to
broadcast artificial noise in order to mask the desired signal from a
potential eavesdropper. The artificial interference is designed to be
orthogonal to the information-bearing signals at the intended
receivers, which ensures that only the eavesdropper suffers a SINR
penalty. Jamming potential eavesdroppers with artificial noise has
been previously proposed for a point-to-point MIMO wiretap channel in
\cite{Negi08,Swindlehurst09}.

For the MIMO broadcast channel with independent signals, we compare the power efficiency and relative SINR of two different approaches: a zero-forcing beamforming design, and an iterative minimum-power joint transmit-receive beamformer design with a minimum SINR constraint per user. The zero-forcing solution allocates slightly lower power for artificial noise at low transmit power levels, but enjoys a significant advantage in terms of complexity. For the MIMO multicast channel, an iterative minimum-power optimal beamformer design is employed with a minimum mean square error (MMSE) criterion for each user.

In the next section, the mathematical models for the MIMO broadcast
and multicast channels are presented. The known algorithms for zero-forcing
beamforming and the optimal minimum-power beamformer design are
outlined in Section~\ref{sec:design}. The wiretapping strategies that
a potential eavesdropper could employ are described in
Section~\ref{sec:wiretapper}. The resulting system performance is
studied via simulation in Section~\ref{sec:sim}, and concluding
remarks are presented in Section~\ref{sec:concl}.

\emph{Notation}: $\mathcal{E}\{\cdot\}$ denotes expectation,
$(\cdot)^T$ the transpose, $(\cdot)^H$ the Hermitian transpose,
$\text{Tr}(\cdot)$ is the trace operator, $\left[ {\mathbf{A}}
\right]_{p,q}$ denotes the $(p,q)$ entry of matrix $\mathbf{A}$,
$\operatorname{diag} \left\{ {\mathbf{x}} \right\}$ is a diagonal
matrix with vector $\mathbf{x}$ on the diagonal, $\left\| \cdot
\right\|_2$ is the Euclidean norm, $\mathbf{1}$ is a column vector of
ones, and $\mathbf{I}$ is an identity matrix of appropriate dimension.

\section{MATHEMATICAL MODEL}\label{sec:model}
The network under consideration is comprised of a $N_t$-antenna
transmitter broadcasting to $K$ legitimate receivers with $N_r$
antennas each, and a single passive eavesdropper with $N_e$ antennas
in the vicinity of the network.  For the moment, we assume $K<N_t$,
and also remark that the proposed secrecy scheme is valid for an
arbitrary number of eavesdroppers. The transmitter is assumed to have
perfect channel state information (CSI) for all of the intended
receivers, but is unaware of either the instantaneous CSI or the
distribution of the CSI for the eavesdropper. This lack of information
precludes the use of secrecy capacity as our performance metric, thus
we choose to work directly with SINR.  The transmitter's
primary objective is to allow each of the desired receivers to recover
the transmitted data with a certain SINR, while denying the
eavesdropper as much information as possible about the data.  Similar
to the approaches taken in \cite{Negi08,Swindlehurst09}, this will be
accomplished by transmission of a jamming signal simultaneously
with the data intended for the desired receivers.

We will treat two separate cases: (1) transmission of a unique data
symbol to each of the $K$ users, referred to as `\emph{broadcasting}',
and (2) sending the same common information signal to all receivers,
referred to as `\emph{multicasting}.' The data model for these
two scenarios is detailed below.

\subsection{MIMO Broadcast Channel}\label{sec:broad}
In the broadcast scenario considered here, the transmitter wishes to
send a private, independent scalar message $z_k$ to each receiver.
The transmitter employs a linear $N_t\times 1$ transmit beamformer
$\mathbf{t}_k$ for the information symbol $z_k$ of each intended
receiver, and is assumed to have a total power constraint $P$
encompassing information transmission and jamming. We denote the
information signal by the $K\times 1$ vector $\mathbf{z}=\left( {z_1 ,
\ldots ,z_K } \right)^T$, and the jamming signal by the $N_t\times 1$
vector $\mathbf{z}'$. Assume unit-norm transmit beamformers
$\mathbf{t}_k^H\mathbf{t}_k=1$, and symbol power
$\mathcal{E}\{|z_k|^2\}=\rho_kP$, where $0< \rho_k \le 1$ is the
fraction of the power devoted to the information signal of user $k$,
and $\rho \equiv \sum\nolimits_{k = 1}^K {\rho _k }$ is the
fraction of the total power used for information transmission.

Define
\begin{equation*}
\mathcal{E}\{\mathbf{z}'\mathbf{z}'^H\} = \mathbf{Q}'_z \qquad \text{Tr}(\mathbf{Q}'_z)= (1-\rho)P,
\end{equation*}
and let ${\mathbf{T}} = \left[ {{\mathbf{t}}_1 {\text{ }} \ldots
{\text{ }}{\mathbf{t}}_K } \right]$ denote the aggregate transmit
beamforming matrix. The signal broadcast by the transmitter is then
given by
\begin{equation}
{\mathbf{x}} = \mathbf{T}{\mathbf{z}} + {\mathbf{z'}}.
\end{equation}
In a flat-fading scenario, the received signal at the $k^{th}$
legitimate receiver, $k=1,\ldots,K$, can be written as
\begin{equation}
{\mathbf{y}}_k  = {\mathbf{H}}_k {\mathbf{t}}_k z_k  + \sum\limits_{j \ne k}^K {{\mathbf{H}}_k {\mathbf{t}}_j z_j }  + {\mathbf{H}}_k {\mathbf{z'}} +{\mathbf{n}}_k,
\end{equation}
where $\mathbf{H}_k$ is the corresponding $N_r\times N_t$ channel
matrix between the transmitter and user $k$, and $\mathbf{n}_k$ is the
naturally occurring i.i.d additive white Gaussian noise vector with
covariance $\mathcal{E}\{\mathbf{n}_b \mathbf{n}_b^H\}= \sigma_n^2
\mathbf{I}$. Analogous parameters can be defined for the eavesdropper,
who receives
\begin{equation}
{\mathbf{y}}_e  =  {\mathbf{H}}_e\sum\limits_{j =1}^K { {\mathbf{t}}_j z_j }  + {\mathbf{H}}_e {\mathbf{z'}} +{\mathbf{n}}_e.
\label{EQ:ye_Eve}
\end{equation}
The $k^{th}$ receiver uses a $N_r\times 1$ beamformer ${\mathbf{w}}_k$
to recover its information, which leads to the decision variable
\begin{equation}
\hat z_k = {\mathbf{w}}_k^H {\mathbf{H}}_k {\mathbf{t}}_k z_k +
{\mathbf{w}}_k^H {\mathbf{H}}_k \sum\limits_{j \ne k}^K
{{\mathbf{t}}_j z_j } + {\mathbf{w}}_k^H {\mathbf{H}}_k {\mathbf{z'}}
+{\mathbf{w}}_k^H {\mathbf{n}}_k.
\end{equation}

\subsection{MIMO Multicast Channel}
In the case of multicast, a common information symbol $z$ with power
$\mathcal{E}\{|z|^2\}=\rho P$ is transmitted to all $K$
receivers. This necessitates the use of a common $N_t\times 1$
transmit beamformer $\mathbf{u}$, with
$\mathbf{u}^H\mathbf{u}=1$. Assuming the same power constraints and
artificial noise properties as in Section~\ref{sec:broad}, the
transmitted signal is
\begin{equation}
{\mathbf{x}} = \mathbf{u}z + {\mathbf{z'}}.
\end{equation}
The received signals are now
\begin{equation}
{\mathbf{y}}_k  = {\mathbf{H}}_k {\mathbf{u}}z  +  {\mathbf{H}}_k {\mathbf{z'}} +{\mathbf{n}}_k
\end{equation}
\begin{equation}
{\mathbf{y}}_e  =  {\mathbf{H}}_e{\mathbf{u}}z  + {\mathbf{H}}_e {\mathbf{z'}} +{\mathbf{n}}_e,
\end{equation}
and the $k^{th}$ receiver employs a $N_r\times 1$ beamformer ${\mathbf{r}}_k$ to obtain its decision variable as
\begin{equation}
\hat z_k  = {\mathbf{r}}_k^H {\mathbf{H}}_k {\mathbf{u}}z  + {\mathbf{r}}_k^H {\mathbf{H}}_k {\mathbf{z'}} +{\mathbf{r}}_k^H {\mathbf{n}}_k.
\end{equation}

\subsection{Artificial Noise}
As will be discussed in the next section, our goal will be to minimize
the total transmit power required to achieve a certain SINR for
each receiver, and use all remaining power to jam the eavesdropper.
To achieve this goal, we will choose the transmit and receive beamformers
in such a way that the jamming signal does not impact the quality of
the desired receiver's signal.  Assuming the transmitter is aware of
the beamformers used by each of the receivers, an effective downlink
channel to the $K$ receivers can be constructed for either the
broadcast or multicast case as follows:
 \begin{equation}
{\mathbf{\tilde H}}  = \left[ {\begin{array}{*{20}c}
   \mathbf{\tilde H}_1 &  \ldots   & \mathbf{\tilde H}_K  \\
 \end{array} } \right]^T  \label{EQ:downlink}
 \end{equation}
where $\mathbf{\tilde H}_k=\left( {{\mathbf{w}}_k^H {\mathbf{H}}_k }
\right)^T$ or $\mathbf{\tilde H}_k=\left( {{\mathbf{r}}_k^H
{\mathbf{H}}_k } \right)^T$.  When $K < N_t$, the jamming signal
$\mathbf{z}'$ can be chosen from the nullspace of $\tilde{\mathbf{H}}$
in order to guarantee that it does not impact the desired receivers.
If $K\geq N_t$, the nullspace of the effective downlink channel does
not exist in general, and the artificial noise could not be guaranteed
to be orthogonal to the desired signals.  Although the artificial
noise can still be constructed so as to minimize its impact at the
receivers ({\em e.g.,} by forcing it to lie in the right subspace of
$\tilde{\mathbf{H}}$ with smallest singular value), a scheduling
strategy to ensure $K< N_t$ may be more appropriate in the context of
this work. We note that user scheduling in wireless networks with
secrecy considerations has received limited attention thus far.

\section{BEAMFORMING DESIGN}\label{sec:design}
In either the broadcast or multicast case, the design of the transmit
and receive beamformers is, in general, coupled; that is, the choice
of $\mathbf{t}_k,\mathbf{u}$ depends on the choice of $\mathbf{w}_k,
\mathbf{r}_k$ respectively, and vice versa.  One solution to this
problem is to fix the beamformer on one end of the link and then optimize
the other.  An optimal approach would design the beamformers jointly,
although at the expense of increased complexity
\cite{Farrokhi02}-\cite{Codreanu07}.  In this paper, we consider both
types of approaches.  In the first, zero-forcing at the transmitter is
used for design of the transmit beamformers in the broadcast case;
this eliminates multi-user interference at each receiver, and leads to
a simple maximum-ratio combiner at each receiver.  In the second
approach, we consider the optimal joint beamformer design problem for
both the broadcast and multicast cases. The lack of eavesdropper CSI preempts the development of beamforming designs tailored towards maximizing a particular secrecy metric; hence we utilize existing precoding methods as discussed in the sequel.

\subsection{Zero-Forcing Beamforming}\label{sec:ZF}
In this section, we adopt a modified version of the coordinated
zero-forcing beamforming approach in \cite{Haardt04}.  Assume
that $K < N_t$, and for user $k$, define ${\tilde{\mathbf{H}}_k}$ as
\begin{equation}
{\mathbf{\tilde H}}_k  = \left[ {\begin{array}{*{20}c}
   {{\mathbf{\tilde h}}_1 } &  \ldots  & {{\mathbf{\tilde h}}_{k - 1} } & {{\mathbf{\tilde h}}_{k + 1} } &  \ldots  & {{\mathbf{\tilde h}}_K }  \\
 \end{array} } \right]  \label{EQ:txbfstep1}
 \end{equation}
 \begin{equation}
  {\mathbf{\tilde h}}_l  = \left( {{\mathbf{w}}_l^H {\mathbf{H}}_l } \right)^T. \nonumber
\end{equation}
The singular value decomposition (SVD) of ${\mathbf{\tilde H}}_k$
yields
\begin{equation}
{\mathbf{\tilde H}}_k  = {\mathbf{\tilde U}}_k {\mathbf{\tilde D}}_k \left[ {{\mathbf{\tilde V}}_k^{\left( s \right)} {\text{ }}{\mathbf{\tilde v}}_k^{\left( 0 \right)} } \right]^H,
\label{EQ:txbf2}
\end{equation}
where $\mathbf{\tilde U}_k$ is the matrix of left singular vectors,
$\mathbf{\tilde D}_k$ is the diagonal matrix of singular values,
${\mathbf{\tilde v}}_k^{\left( 0 \right)}$ is the right singular vector
associated with the smallest (zero) singular value, and
${\mathbf{\tilde V}}_k^{\left( s \right)}$ is the collection of right
singular vectors corresponding to other singular values.

Evidently, ${\mathbf{v}}_k^{\left( 0 \right)}$ is a logical choice for
the $k^{th}$ transmit beamformer $\mathbf{t}_k$ given the objective of
nulling all multiuser interference.  Since receiver $k$ then sees only
its desired signal in spatially white noise, the optimal receive
beamformer ${\mathbf{w}}_k$ is simply the maximum-ratio combiner:
\begin{equation}
{\mathbf{w}}_k  = {\mathbf{H}}_k {\mathbf{t}}_k.
\end{equation}
Assuming that the jamming signal is orthogonal to $\mathbf{w}_k^H \mathbf{H}_k$,
the SINR at user $k$ can then be written as
\begin{equation}
\mbox{\rm SINR}_k = \frac{\rho_k P | \mathbf{w}_k^H \mathbf{H}_{k} \mathbf{t}_k |^2}{
\sigma_n^2 \mathbf{w}_k^H \mathbf{w}_k}.
\end{equation}
For a target SINR $S_k$, the required power allocation for user $k$ can be calculated as
\begin{equation}
\rho_k = \frac{\sigma_n^2 S_k}{\mathbf{t}_k^H\mathbf{H}_{k}^H\mathbf{H}_{k}\mathbf{t}_k P}.
\end{equation}


\subsection{Optimal Minimum Transmit Power Beamforming with per-user SINR constraint}\label{sec:iter}
Although relatively simple, the proposed zero-forcing algorithm in
Section~\ref{sec:ZF} will not in general minimize $\rho P$.  To
minimize the transmit power necessary to achieve the desired SINR, it
is necessary to jointly design the transmit and receive beamformers
\cite{Farrokhi02}-\cite{Codreanu07}.  Since the
optimal beamformers will not be of the zero-forcing type,
the downlink beamformer design problem is non-convex due to the
interdependence of the problem variables. This issue can be overcome
by exploiting the SINR duality of the downlink and uplink channels,
which states that the minimum sum power required to achieve a set of
SINR values on the downlink is equal to the minimum sum power required
to achieve the same SINR vector on the dual uplink channel
\cite{Codreanu07}. Therefore, as a benchmark we compute the optimum
transmit/receive beamformers and power allocation that minimizes the
sum transmit power while satisfying the SINR constraint per user based
on \cite{Spencer04,Codreanu07}.

Let ${{\mathbf{t}}_k^{\left( i \right)} }$, ${{\mathbf{w}}_k^{\left( i
\right)} }$, $p_k^{\left( i \right)}$ and $q_k^{\left( i \right)}$
represent the transmit/receive beamformers and downlink/uplink power
allocation for user $k$ at the $i^{th}$ iteration.  Define $g_{s,k}$ for
$s,k=1,\ldots,K$ as the signal and interference powers for each user:
\begin{equation}
g_{s,k}  \triangleq \left\{ {\begin{array}{*{20}c}
   {\left| {\left( {{\mathbf{t}}_s^{\left( i \right)} } \right)^H {\mathbf{H}}_{k,s}^H {\mathbf{w}}_k^{\left( i \right)} } \right|^2 }  \text{for uplink}\\
   {\left| {\left( {{\mathbf{w}}_k^{\left( i \right)} } \right)^H {\mathbf{H}}_{k,s}^{} {\mathbf{t}}_s^{\left( {i + 1} \right)} } \right|^2 }  \text{for downlink}\\

 \end{array} } \right.
\end{equation}
The SINR per stream on either link can then be written as
\begin{equation}
\gamma _k  = \frac{{x_k \left| {g_{s,s} } \right|^2 }}
{{1 + \sum\nolimits_{i = 1,i \ne s}^K {x_i \left| {g_{s,i} } \right|^2 } }}
\label{eq:sinr}
\end{equation}
where $x_k$ is the power allotted to user $k$ either on the downlink or the dual uplink channel.
Finally, define matrices $\mathbf{C}$ and $\mathbf{D}$ as
\begin{equation}
\left[ {\mathbf{C}} \right]_{s,k}  = \left\{ {\begin{array}{*{20}c}
   {g_{s,k} } & {{\text{if }}s \ne k}  \\
   0 & {{\text{if }}s = k}  \\

 \end{array} } \right.
 \end{equation}

\begin{equation}
{\mathbf{D}} = \operatorname{diag} \left\{ {\frac{{\gamma _{th} }}
{{g_{1,1} }}, \ldots ,\frac{{\gamma _{th} }}
{{g_{K,K} }}} \right\}.
\end{equation}

The iterative beamformer design that minimizes the transmit power can be summarized as follows.
\begin{enumerate}
\item Set the initial transmit beamformers to the zero-forcing solution ${\mathbf{t}}_k^{\left( 0 \right)}  = {\mathbf{\tilde v}}_k^{\left( 0 \right)}$, with an initial power allocation $p_k^{\left( 0 \right)}=P/K$ per user. Compute the optimum set of receive beamformers as
    \begin{equation}
{\mathbf{w}}_k^{\left( i \right)}  = \sqrt {p_k^{\left( i \right)} } {\mathbf{t}}_k^{\left( i \right)H} {\mathbf{H}}_k^H \left( {\sum\limits_{i = 1}^K {p_k^{\left( i \right)} {\mathbf{H}}_k {\mathbf{t}}_k^{\left( i \right)} {\mathbf{t}}_k^{\left( i \right)H} {\mathbf{H}}_k^H  + {\mathbf{I}}} } \right)^{ - 1}_. \label{eq:rxupdate}
\end{equation}
\item Now consider the dual uplink channel where the transmit beamformers from Step 1 serve as the receive beamformers, and vice versa. Update the power allocation vector $\mathbf{q}^{\left( i +1\right)}=\mathbf{x}^*$ from
\begin{equation}
{\mathbf{x}}^*  = \left( {{\mathbf{I}} - {\mathbf{DC}}} \right)^{ - 1} {\mathbf{D1}},  \label{eq:power}
\end{equation}
then update the beamformer set using (\ref{eq:txupdate}):
\begin{equation}
{\mathbf{t}}_k^{\left( i \right)}  = \sqrt {q_k^{\left( i \right)} } {\mathbf{w}}_k^{\left( i \right)H} {\mathbf{H}}_k \left( {\sum\limits_{i = 1}^K {q_k^{\left( i \right)} {\mathbf{H}}_k^H {\mathbf{w}}_k^{\left( i \right)} {\mathbf{w}}_k^{\left( i \right)H} {\mathbf{H}}_k  + {\mathbf{I}}} } \right)^{ - 1}_.  \label{eq:txupdate}
\end{equation}
Calculate the achieved SINR set on the uplink using (\ref{eq:sinr}), and go to Step 3.
\item Revert back to the downlink channel and recompute the power allocation vector $\mathbf{p}^{\left( i +1\right)}=\mathbf{x}^*$ using (\ref{eq:power}). Calculate the downlink SINR set using (\ref{eq:sinr}) and compare with the uplink SINR vector from Step 2 for convergence. If the stopping criterion is not satisfied, set $i=i+1$ and return to Step 2, otherwise terminate the algorithm.
\end{enumerate}

\subsection{Optimal Minimum Transmit Power Beamforming with per-user MSE constraint}\label{sec:MMSEbf}
Despite superficial similarities, the beamforming design problems with
per-user SINR constraints for the broadcast and multicast channels are
fundamentally different. Many efficient numerical solutions exist for
the former as cited in Section~\ref{sec:ZF}, whereas the multicast
problem is known to be non-convex even for the MISO case with single
antennas at each receiver. A number of approximate solutions based on
semidefinite relaxation techniques have been proposed for the MISO
multicast channel, e.g. \cite{Sidiropoulos06,Sidiropoulos08}.

However, the MIMO multicast beamforming problem was recently reformulated as a convex optimization by replacing the per-user SINR requirements with minimum MSE constraints \cite{Boche08}. In this case, the optimal receiver structure is known to be MMSE, which allows an alternating iterative optimization of the transmit and receive beamformers. The minimum sum-power optimization problem can be expressed as a convex second-order cone program (SOCP):
\begin{equation}
\begin{gathered}
  \mathop {\min }\limits_{{\mathbf{u}},t} {\text{ }}t \hfill \\
  {\text{s}}{\text{.t}}{\text{. }}\left\| {\mathbf{u}} \right\|_2  \leqslant t, \hfill \\
  \left\| {v_k } \right\|_2  \leqslant \sqrt {\varepsilon _k  - w_k } ,{\text{ }}\forall k \hfill \\
\end{gathered} \label{eq:SOCP}
\end{equation}
where $\varepsilon_k$ is the MMSE constraint per user, $v_k =
{\mathbf{r}}_k^H {\mathbf{H}}_k {\mathbf{u}} - 1$, and $w_k = \sigma
_n^2 \operatorname{Tr} \left( {{\mathbf{r}}_k {\mathbf{r}}_k^H }
\right)$.

In brief, the algorithm is initialized with random receive
beamformers, after which the corresponding optimal transmit beamformer is
computed by solving (\ref{eq:SOCP}). The receive beamformers are
updated using the MMSE criterion as
\begin{equation}
{\mathbf{r}}_k  = {\mathbf{u}}^H {\mathbf{H}}_k^H \left( {{\mathbf{H}}_k {\mathbf{uu}}^H {\mathbf{H}}_k^H  + \sigma _n^2 {\mathbf{I}}} \right)^{ - 1},
\end{equation}
and the iterations continue until
convergence. For consistency with our choice of SINR as the
performance metric, the following equivalence relation between maximum
SINR and MMSE is useful:
\begin{equation}
\varepsilon _k  = \frac{1}
{{1 + \operatorname{SINR} _k }}.
\end{equation}

\emph{Remark 1}: For both zero-forcing and joint minimum-power
beamformer designs, the computations are carried out at the
transmitter, which then needs to supply the receivers with information
about the optimal beamformers. This can be done using a limited
(quantized) feedforward scheme, as proposed in \cite{Heath08}.

\emph{Remark 2}: The assumption of perfect CSIT of the intended receivers is admittedly a strong one. Robust beamforming design for the MIMO downlink with multi-antenna receivers is an ongoing research problem, with some recent results provided in \cite{Wong08}. However, incorporating artificial noise into any such robust beamforming schemes is not a straightforward extension, since the lack of exact receiver CSI at the transmitter would no longer allow any artificial noise to be perfectly orthogonal to the intended receivers. The authors have conducted a perturbation analysis to capture the performance degradation in single-user MIMO wiretap channels \cite{TSP09}, and an extension to the downlink case is in progress.

\emph{Remark 3}: As mentioned previously, the assumption that the number of receivers is less than the number of transmit antennas in Section II-C can be relaxed by implementing a user selection stage prior to transmission. The authors have shown in \cite{Asilomar09} that a greedy algorithm which schedules the user set based on maximizing the transmit power available for artificial noise performs close to an optimal exhaustive search in terms of eavesdropper BER.

\section{WIRETAPPING STRATEGIES}\label{sec:wiretapper}
We consider two types of eavesdropper strategies: (1) a simple linear
receiver approach in which the eavesdropper attempts to maximize the
SINR of the data stream she is interested in decoding, and (2) a
multi-user decoding scheme in which the eavesdropper uses maximum
likelihood detection to decode all $K$ information-bearing waveforms.
We begin by illustrating approach~(1) for the broadcast case, with the
extension to the multicast channel being straightforward. Assume that
the eavesdropper seeks to recover the data stream of user $k$ from her
received signal given in (\ref{EQ:ye_Eve}). The
interference-plus-noise covariance matrix given that $z_k$ is the
symbol of interest is
\begin{equation}
{\mathbf{Q}}_e^k  = {\mathbf{H}}_e \sum\limits_{j \ne k}^K {{\mathbf{t}}_j {\mathbf{t}}_j^H {\mathbf{H}}_e^H }  + {\mathbf{H}}_e {\mathbf{Q}}'_z {\mathbf{H}}_e^H  + \sigma _e^2 {\mathbf{I}}.
\end{equation}
The maximum SINR beamformer for the data stream of user $k$ is then given by
\begin{equation}
\mathbf{w}_e = \left(\mathbf{Q}_e^k\right)^{-1}\mathbf{H}_{e} \mathbf{t}_k.
\end{equation}
The use of an optimal beamformer here presumes that
$\mathbf{H}_{e}\mathbf{t}_k$, $k=1,\ldots,K$ is somehow known at the
eavesdropper.  Using this approach, the SINR at the eavesdropper
can be calculated to be
\begin{equation}
\operatorname{SINR}_{e,k} = \mathbf{t}_k^H \mathbf{H}_e^H \left(\mathbf{Q}_e^k\right)^{-1}
\mathbf{H}_{e} \mathbf{t}_k.
\end{equation}

An eavesdropper with more extensive computational resources
will attack the MIMO broadcast network using a more sophisticated
approach.  In general, the optimal decoder at the eavesdropper
in terms of minimizing the symbol error rate would be the
following Maximum Likelihood (ML) detector:
\begin{equation}
{\mathbf{\hat z}} = \arg \mathop {\min }\limits_{{\mathbf{z}} \in \mathcal{Z}} \left\{ {\left\| {\mathbf{Q}}_e^{ - 1/2}\left({{\mathbf{y}}_e  - {\mathbf{H}}_e \mathbf{T} {\mathbf{z}}}\right) \right\|^2 } \right\}
\end{equation}
where $\mathcal{Z}$ is the signal space from which $\mathbf{z}$ is
drawn, and $\mathbf{Q}_e={\mathbf{H}}_e {\mathbf{Q}}'_z
{\mathbf{H}}_e^H + \sigma _e^2 {\mathbf{I}}$ is the
interference-plus-noise covariance matrix perceived by the
eavesdropper.

In the next section, we present numerical examples that show
the SINR and the BER that the eavesdropper experiences
with the proposed jamming scheme.

\section{SIMULATION RESULTS}\label{sec:sim}
We investigate the performance of the eavesdropper and the desired
receivers as a function of the target SINR at the receivers and the
total available transmit power. In each simulation, we assume a
transmitter with $N_t=4$ antennas, $K=3$ legitimate receivers with
$N_r=2$ antennas each, and an eavesdropper with $N_e=4$ antennas. The
channel matrices for all links are composed of independent Gaussian
random variables with zero mean and unit variance. The background
noise power is assumed to be the same for all $K$ receivers and the
eavesdropper: $\sigma_n^2=\sigma_e^2=1$.  The algorithm of
\cite{Spencer04} is used to implement the optimal joint beamformer
design, and the SOCP in (\ref{eq:SOCP})
was solved using the MATLAB $\operatorname{cvx}$ optimization toolbox
\cite{Boyd}.  All of the displayed results are calculated based on an
average of 5000 trials with independent channel and noise
realizations.

\begin{figure}[htbp]
\centering
\includegraphics[width=3.5in]{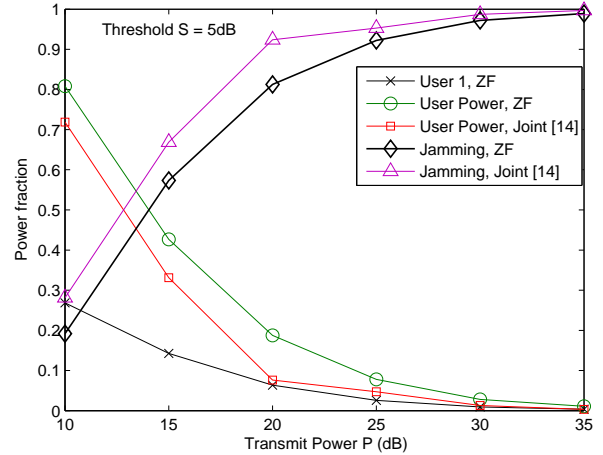}
\caption{Power fraction versus transmit power $P$ for $K =3$ users, $N_t=4, N_r=2, N_e=4$ antennas.}
\label{fig_rho}
\end{figure}
Figure~\ref{fig_rho} displays the average fraction of the transmit
power allocated to data and artificial noise by the zero-forcing and
the optimal minimum transmit power beamforming algorithms with an SINR
threshold of 5~dB. The joint design requires roughly $10\%$ less
transmit power at small to intermediate power levels, albeit with a
significantly greater level of complexity.  The curve labeled ``User
1,ZF'' represents the fraction of the total power assigned to an
arbitrary user (referred to as user~1) among the three legitimate
receivers.

\begin{figure}[htbp]
\centering
\includegraphics[width=3.5in]{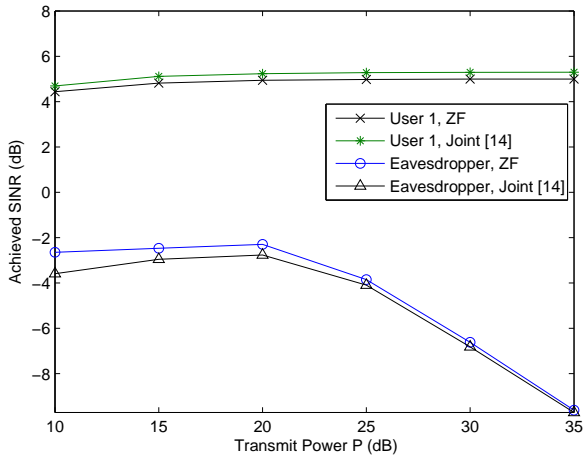}
\caption{SINR versus transmit power $P$ for $K =3$ users, $N_t=4, N_r=2, N_e=4$ antennas.}
\label{fig_SINR}
\end{figure}

Figure~\ref{fig_SINR} shows the achieved SINR levels versus transmit
power $P$ for user~1 and the average SINR averaged over all three
streams for the eavesdropper when using single-user detection. The
legitimate receiver almost always achieves the desired SINR target of
5 dB, while the eavesdropper has a significantly lower SINR due to the
artificial noise.  The SINR of user~1 is slightly below 5~dB for
transmit powers of 10 and 15 dB, since there were a few trials for
which the 5~dB SINR target could not be achieved.  In such cases, the
transmitter devotes all power to the desired receivers and none to
jamming, and the resulting SINR is averaged in with the other trials.
Note that there is not a significant difference in performance for the
eavesdropper whether the zero-forcing or optimal broadcast beamformers
are used.

\begin{figure}[htbp]
\centering
\includegraphics[width=3.5in]{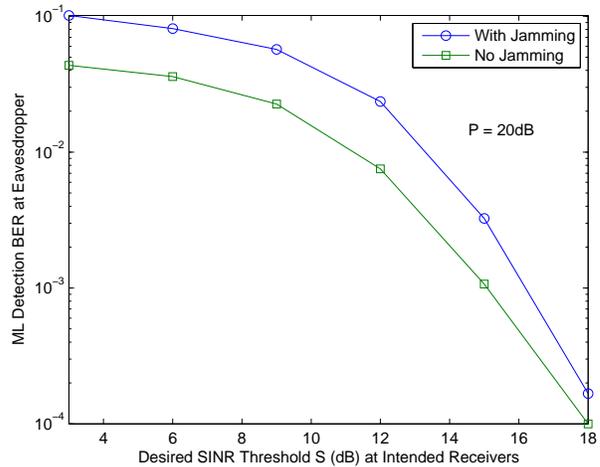}
\caption{Eavesdropper maximum-likelihood BER versus target SINR $S$ for $K =3$ users, $N_t=4, N_r=2, N_e=4$ antennas.}
\label{fig_BER}
\end{figure}
Figure~\ref{fig_BER} compares the eavesdropper's BER with and without
artificial noise when the eavesdropper employs MIMO ML detection, assuming an uncoded BPSK-modulated information signal
$\mathbf{z}$ and zero-forcing transmit beamforming. For low target SINRs, we observe a significant
degradation in the eavesdropper's interception capabilities, e.g., by approximately
10.5 dB at $BER=0.05$. A more modest gain of 2.5 dB is achieved at $BER=10^{-2}$ for intermediate target SINRs. At high SINR thresholds, the two
curves converge since the transmitter is constrained to allocate
progressively smaller fractions of the total power to jamming.

\begin{figure}[htbp]
\centering
\includegraphics[width=3.5in]{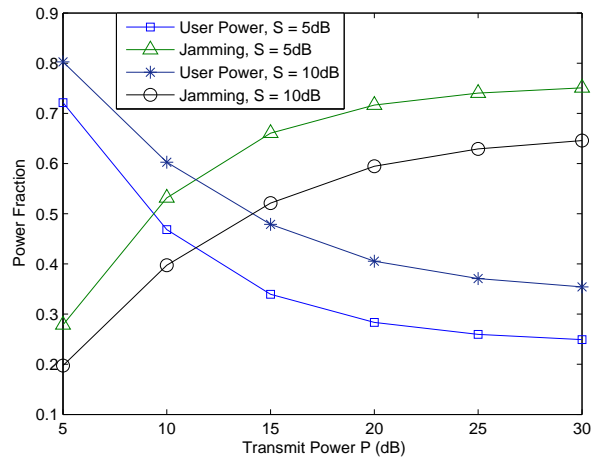}
\caption{Power fraction versus transmit power $P$ for $K =3$ users, $N_t=4, N_r=2, N_e=4$ antennas.}
\label{fig_Multicastrho}
\end{figure}
Figure~\ref{fig_Multicastrho} displays the fraction of the transmit
power allocated to data and artificial noise by the multicast minimum
transmit-power beamforming algorithm of Section~\ref{sec:MMSEbf}, with
SINR thresholds of $S=5$ dB, 10 dB. The lack of inter-user
interference allows the transmitter to allocate more power for jamming
compared to the broadcast case for the same transmit power.

\begin{figure}[htbp]
\centering
\includegraphics[width=3.5in]{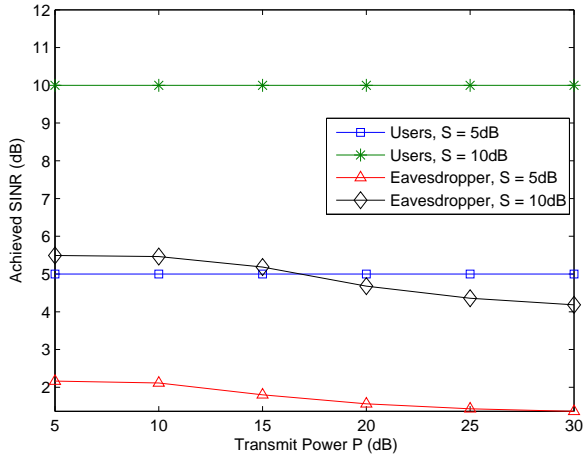}
\caption{SINR versus transmit power $P$ for $K =3$ users, $N_t=4, N_r=2, N_e=4$ antennas.}
\label{fig_MulticastSINR}
\end{figure}
Figure~\ref{fig_MulticastSINR} shows the average achieved SINR
levels at the intended receivers and the eavesdropper versus transmit
power $P$ for user SINR thresholds of $S=5$ dB, 10 dB. As before, the
legitimate receivers achieve the desired SINR targets, while the
eavesdropper's performance is degraded. However, the degradation in
the eavesdropper's SINR is not as severe as in the broadcast case since
there is no multiuser interference to compound the effect of the
artificial noise.

\section{CONCLUSION}\label{sec:concl}
This paper has examined beamforming strategies combined with artificial
noise for providing confidentiality at the physical layer in multiuser
MIMO wiretap channels. For the MIMO broadcast channel with single-user
detection at the eavesdropper, the zero-forcing beamformer design is
shown to provide an acceptable level of performance in terms of
relative SINR when compared to optimal joint transmit-receive
beamforming algorithms. The use of artificial noise is meaningful even
when the eavesdropper employs optimal ML joint detection for the
information vector. For the MIMO multicast channel, the degradation in
the eavesdropper's SINR as the transmit power increases is not as
severe, but artificial noise is still seen to be effective.


\end{document}